\documentclass[12pt]{article}
\usepackage{psfig}

\def\lesssim{\mathrel{\mathpalette\vereq<}}
\def\gtrsim{\mathrel{\mathpalette\vereq>}}
\makeatletter
\def\vereq#1#2{\lower3pt\vbox{\baselineskip1.5pt \lineskip1.5pt
\ialign{$\m@th#1\hfill##\hfil$\crcr#2\crcr\sim\crcr}}}
\makeatother

\begin{document}
\begin{titlepage}
\begin{center}
\hfill    CERN-TH/2000-067\\
\hfill LBNL-45322\\
\hfill UCB-PTH-00/06\\
~{} \hfill hep-ph/0003210\\

\vskip .1in

{\large \bf Determining the Flavour Content\\of the Low-Energy Solar Neutrino 
Flux}

\vskip 0.3in

Andr\'e de Gouv\^ea

\vskip 0.05in

{\em CERN - Theory Division\\
     CH-1211 Geneva 23, Switzerland}

\vskip 0.1in

Hitoshi Murayama

\vskip 0.05in

{\em Theoretical Physics Group, Lawrence Berkeley National Laboratory \\ 
      Berkeley, California 94720 \\ 
      {\rm and}\\
      Department of Physics, University of California \\  
      Berkeley, California 94720}

\end{center}

\vskip .1in

\begin{abstract}
We study the sensitivity of the HELLAZ and Borexino solar neutrino
experiments on discriminating the neutrino species $\nu_e$,
$\bar{\nu}_e$, $\nu_{\mu,\tau}$, $\bar{\nu}_{\mu,\tau}$, and $\nu_{s}$ using the
difference in the recoil electron kinetic energy spectra in 
elastic neutrino-electron
scattering.  We find that one can observe a non-vanishing
$\nu_{\mu,\tau}$ component in the solar neutrino flux, especially when
the $\nu_e$ survival probability is low.  Also, if the data turn out to be 
consistent with $\nu_{e} \leftrightarrow \nu_{\mu,\tau}$ 
oscillations, a $\bar{\nu}_e$ component can be excluded effectively. 
\end{abstract}

\end{titlepage}

\newpage
\setcounter{footnote}{0}
\section{Introduction\label{sec:introduction}}

The flux of solar neutrinos was first measured in the Homestake mine 
(see \cite{Cl} and references therein) over thirty years ago. Since then, it
was realized that the measured flux was significantly suppressed with respect
to theoretical predictions. More recently, a handful of different experiments
have also succeeded in measuring the solar neutrino flux 
\cite{Kamiokande, GALLEX,
SAGE, Super-K}. All experiments measure a neutrino flux which is significantly
suppressed with respect to the theoretical predictions of the most recent
version of the Standard Solar Model (SSM) \cite{SSM}. This thirty year old
problem is what is referred to as the ``solar neutrino puzzle.''

There are different types of solutions to the solar neutrino
puzzle. At first sight,
it appears natural to suspect that the SSM predictions for the solar
neutrino flux 
are slightly off, and/or that the experiments have underestimated their 
systematic effects, given that detailed models of the Sun and neutrino 
experiments are highly non-trivial. However, SSM independent analyses of 
the neutrino data (see \cite{ssm_indep} for a particularly nice and simple
example), together with independent experimental evidence in favour of the
SSM \cite{SSM}, seem to indicate that the above solution to the puzzle is
strongly disfavoured.

The best solution to the solar neutrino puzzle involves extending the
Standard Model of particle physics by assuming that the neutrinos have mass
and that they mix, {\it i.e.,}\/ neutrino mass eigenstates are different from 
neutrino weak eigenstates. This possibility has become particularly natural
in light of the recent strong evidence for $\nu_{\mu}$ oscillations from the 
atmospheric neutrino data at SuperKamiokande \cite{atmospheric}. 

Nonetheless, in order to firmly establish that the solution to the solar 
neutrino puzzle involves physics beyond the Standard Model, it is necessary
to come up with SSM independent, robust experimental evidence for, 
{\it e.g.}\/, solar neutrino oscillations. Indeed, these ``Smoking Gun'' 
signatures of solar $\nu_e\leftrightarrow\nu_{\rm other}$ oscillations 
are among the present goals of the SuperKamiokande experiment, 
via the measurement of the 
day-night asymmetry of the solar neutrino data \cite{SK_daynight}
and the recoil electron energy spectrum \cite{SK_spectrum}, 
and the SNO experiment \cite{SNO}, via the measurement of the 
charged to neutral current ratio, the day-night asymmetry of the data, and
the recoil electron kinetic energy spectrum.

Other goals of this and the next generation of neutrino experiments
are, if solar neutrino oscillations are established, to determine
neutrino oscillation modes and measure masses and mixing angles. The
current data allow for different $\nu_e$ oscillation modes and a
handful of disconnected regions in the mass--mixing-angle parameter
space (see \cite{bksreview,rate_analysis} for two-flavour analyses and
also \cite{darkside} for an extension to the ``dark side'' of the
parameter space).

Experiments dedicated to measuring the flux of low-energy solar neutrinos 
($E_{\nu}=O(100-1000)$~keV) are going to be extremely useful, and perhaps 
crucial, in order to fully solve the solar neutrino puzzle. It was recently
shown that future experiments (Borexino \cite{Borexino} and, perhaps, 
KamLAND \cite{KamLAND}) dedicated to measuring the flux of $^7$Be 
neutrinos (produced by $^7$Be+$e^{-}\rightarrow^7$Li$+\nu_e$ inside the Sun) 
should be able to establish or exclude the ``just-so'' solution 
\cite{bksreview} to the solar
neutrino puzzle via the study of the seasonal variations of the neutrino flux
\cite{seasonal}, and establish or exclude the LOW MSW solution 
\cite{rate_analysis} via the study of the zenith angle dependence of their 
data \cite{daynight}. Furthermore, the measurement of a sizable
$^7$Be neutrino flux would significantly disfavour the SMA MSW 
\cite{rate_analysis} solution to the
solar neutrino puzzle, especially in the case of $\nu_e\leftrightarrow\nu_{s}$
oscillations (where $\nu_s$ is a sterile neutrino, {\it i.e.,}\/ a 
standard model singlet), and 
significantly constrain the SSM independent analysis, which require the
flux of $^7$Be neutrinos to be virtually absent \cite{ssm_indep}.  
Finally, we have shown \cite{recoil_spectrum} that, in the advent that
the background rates at Borexino and/or KamLAND are exceptionally low, it 
should be possible to measure a nonzero component of $\nu_{\mu,\tau}$ in the
solar neutrino flux by analysing the recoil electron kinetic energy spectrum. 

Another exciting possibility is that of measuring the ``fundamental''
$pp$-neutrinos, which are produced in the interior of the Sun by
proton-proton fusion ($p+p\rightarrow ^2$H$+e^++\nu_e$) in a real time
experiment. Future experiments, such as HELLAZ, HERON,
LENS, etc (see \cite{Lanou} for an overview) are being designed to do
just that.  The flux of $pp$-neutrinos is particularly constrained by
the photon flux {\it i.e.,}\/ the Sun's luminosity, which, of course,
is very well measured on the Earth. These lowest energy solar
neutrinos ($E_{\nu}\lesssim420$~keV) are not only the most abundant
ones, but also have the best known flux.  Their energy spectrum is
also very well known, since it is dictated by the particularly well
studied $p+p$ nuclear fusion reaction.  Among these proposed 
experiments, HELLAZ \cite{HELLAZ} will be able to determine the
incoming neutrino  
energy in an event-by-event basis and have the unique opportunity of 
studying the solar neutrino spectrum and the recoil electron 
kinetic energy spectrum separately.
Similar to what was shown for $^7$Be neutrinos \cite{recoil_spectrum}, the 
authors of \cite{spectrum_first} showed that HELLAZ may be able to measure 
a nonzero component of $\nu_{\mu,\tau}$ in the solar $pp$-neutrino flux 
by analysing the recoil electron kinetic energy spectrum independent of 
the SSM prediction for the solar neutrino flux. 

In this paper we extend the analysis done in \cite{recoil_spectrum}, and 
study the flavour composition of the flux of $pp$ and $^7$Be neutrinos using
the recoil electron kinetic energy spectrum. In particular we will address
the capability of future low-energy solar-neutrino experiments to see
evidence for $\nu_{\mu,\tau}$ coming from the Sun, and, in light of 
such evidence, exclude more ``exotic'' oscillation scenarios, such as 
$\nu_e\leftrightarrow\nu_s$ or $\nu_e\leftrightarrow\bar{\nu}_{\rm any}$ 
oscillations. 

Our presentation is organised as follows: Sec.~2 describes the flavour
dependent recoil kinetic energy distribution of events at Borexino and HELLAZ. 
Sec.~3 presents the technique for determining the presence of 
$\nu_{\mu,\tau}$ coming from the Sun, independent of the SSM prediction for
the neutrino flux. We present simulations for both Borexino and HELLAZ and
show how such a determination can be improved once we take the SSM prediction
for the neutrino flux into account. Sec.~4 describes how the same
procedure can be used to exclude the presence of antineutrinos or 
sterile neutrinos in the solar neutrino flux. In Sec.~5, we conclude.  

\setcounter{footnote}{0}
\section{Recoil Electron Kinetic Energy Spectrum\label{sec:recoil}}
 
In this section, we discuss the differences in the recoil electron
kinetic energy spectra among different neutrino species.  Low-energy
solar neutrinos are detected via ``$\nu$''$+e^-\rightarrow$``$\nu$''
$+e^-$ elastic scattering in the experiments which will be considered
here. By ``$\nu$'' in the previous sentence, one actually means any of
$\nu_{e}$, $\nu_{\mu}$, $\nu_{\tau}$, $\bar{\nu}_{e}$,
$\bar{\nu}_{\mu}$, or $\bar{\nu}_{\tau}$. Because $\nu_{\mu}$ and
$\nu_{\tau}$ are indistinguishable as far as the reaction above is
concerned, we will refer to both as $\nu_{\mu}$.

\begin{table}
\caption{Coefficients $A$, $B$ in Eq.~\protect\ref{eq_dist} for
  different neutrino species.}
\begin{center}
\begin{tabular}{|c|c|c|}
\hline
species & $A$ & $B$
 \\ \hline
$\nu_e$ & $\sin^2 \theta_W + \frac{1}{2}$ & $\sin^2 \theta_W$ \\ 
\hline
$\bar{\nu}_e$ & $\sin^2 \theta_W$ & $\sin^2 \theta_W + \frac{1}{2}$ \\ 
\hline
$\nu_{\mu,\tau}$ & $\sin^2 \theta_W - \frac{1}{2}$ & $\sin^2 \theta_W$
\\ 
\hline
$\bar{\nu}_{\mu,\tau}$ & $\sin^2 \theta_W$ & $\sin^2 \theta_W - \frac{1}{2}$ \\
\hline
\end{tabular}
\end{center}
\label{AB}
\end{table}

The kinetic energy distribution of the recoil electrons, for a given incoming
neutrino energy $E_{\nu}$ is very well known and given by \cite{tHooft}
\begin{equation}
\frac{{\rm d}\sigma(T,E_{\nu})}{{\rm d}T}=\frac{2G_F^2m_e}{\pi}\left[A^2+B^2
\left( 
1-\frac{T}{E_{\nu}}\right)^2-AB\frac{m_eT}{(E_{\nu})^2} 
\right],
\label{eq_dist}
\end{equation}
where $m_e$ is the electron mass, $T$ is the kinetic energy of the recoil
electron, and $G_F$ is the Fermi constant.
The parameters $A$ and $B$ are given in Table~\ref{AB}.  The
sign difference in the term $1/2$ is a consequence of the presence
(absence) of $W$-boson exchange and the interchange of $A$ and $B$
between neutrino and anti-neutrino cases is a consequence of the
``handedness'' of the weak interactions.
Eq.~(\ref{eq_dist}) is a tree-level expression, but higher order corrections
are known to be very small \cite{corrections}, especially for the neutrino
energies of interest, and will be neglected throughout. 

Borexino (under construction) is an ultra-pure liquid scintillator
tank which detects  
the scintillating light produced by the recoil electron absorbed by the 
medium. For more details see \cite{Borexino, seasonal}. It is sensitive to 
recoil electron kinetic energies greater than 250~keV, and is therefore 
sensitive to the (almost) monochromatic $^7$Be neutrinos with $E_{\nu}=862$~keV.
The expected resolution for the kinetic energy measurement varies
from roughly 12\%, for $T=T_{\rm min}=0.25$~MeV, to 7\% for 
$T=T_{\rm max}=0.66$~MeV \cite{Borexino}.
They expect 53 events/day in the SSM (BP95)
together with 19 background events/day with
the anticipated radiopurity of the scintillator of $10^{-16}$g/g for
U/Th, $10^{-18}$g/g for $^{40}$K, and $^{14}$C$/{}^{12}$C$ = 10^{-18}$ and
no radon diffusion.  It is remarkable, however, that the M\"unchen
group of Borexino achieved a radiopurity for an organic liquid
(Phenyl-ortho-xylylethane) better than $1.0 \times
10^{-17}$g/g \cite{1E-17}; this is an upper bound on the contamination,
limited by the sensitivity of the neutron activation measurement and
hence the actual radiopurity may be even better.  In this paper, we
ignore the background to the $^7$Be solar neutrino signal at Borexino.
This is probably an overoptimistic assumption, but could be realised in
future upgrades given the above-mentioned achievement.

HELLAZ (proposed) 
is a large time projection chamber (TPC) filled with roughly 2000~m$^3$ of 
cool helium gas ($\sim 6$ tons at 5 atmos, 77~K), 
which serves as the target for $\nu$-$e$ scattering. The recoil electron 
propagates in the gas medium before being absorbed, leaving a track of ionization
electrons. These are then collected, yielding information about the 
kinetic energy and the flight direction of the recoil electron.   
HELLAZ is sensitive to recoil kinetic energies greater than $\sim50$~keV,
and can therefore ``see'' most of the $pp$-neutrino spectrum.  
Most importantly,
since not only the recoil kinetic energy of scattered electrons is measured but
also their direction, it is possible to reconstruct the incoming neutrino energy,
given that the position of the Sun in the sky is known, 
via the simple kinematic relation
\begin{equation}
T=m_e\frac{2\cos^2\theta}{(1+m_e/E_{\nu})^2-\cos^2\theta},
\label{T_theta}
\end{equation} 
where $\theta$ is the recoil electron scattering angle with respect to the 
incoming neutrino direction in the laboratory frame.
Incidently, from Eq.~(\ref{T_theta}) it is very easy to compute the 
maximum value of the recoil electron kinetic energy, 
$T_{\rm max}=T(\theta=0)=E_{\nu}/(1+m_e/(2E_{\nu}))$.
HELLAZ expects to measure the recoil electron 
kinetic energy with a resolution which varies
roughly from 2\% to 4\% and the incoming neutrino energy with a resolution
which varies between 5\% and 12\% \cite{HELLAZ}.  
They expect around 7 events/day from $pp$ neutrinos 
in the SSM with negligible background.  
The major sources of background  
at HELLAZ are radioactive impurities from $^{232}$Th and $^{238}$U in
the structure 
of the TPC. However, because of the detector's total event reconstruction 
capabilities (including directional information), very efficient background
rejection schemes are possible (see \cite{HELLAZ} and references therein
for further information).

\begin{figure} [h]
\centerline{
  \psfig{file=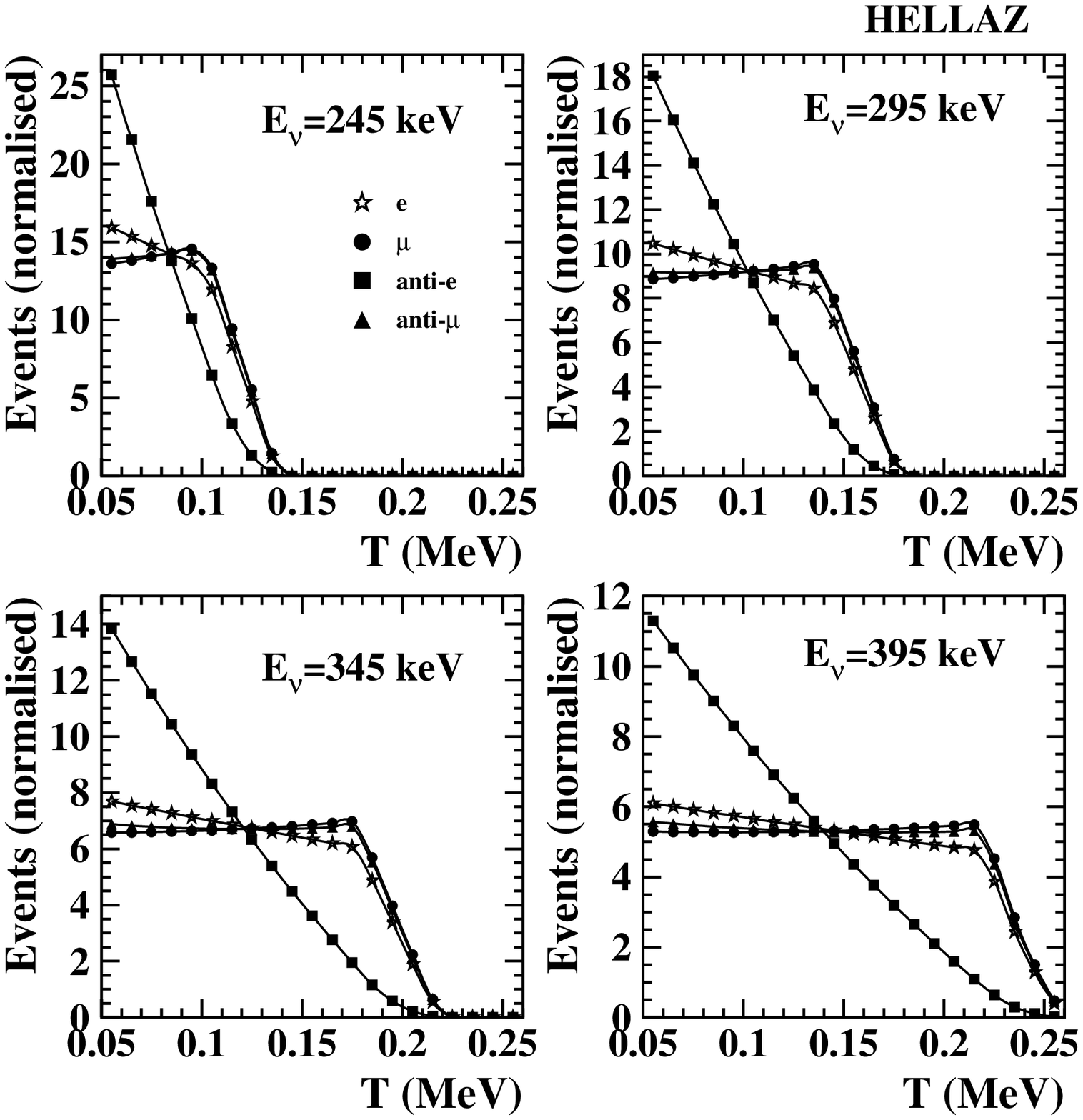,width=0.7\textwidth}
  \psfig{file=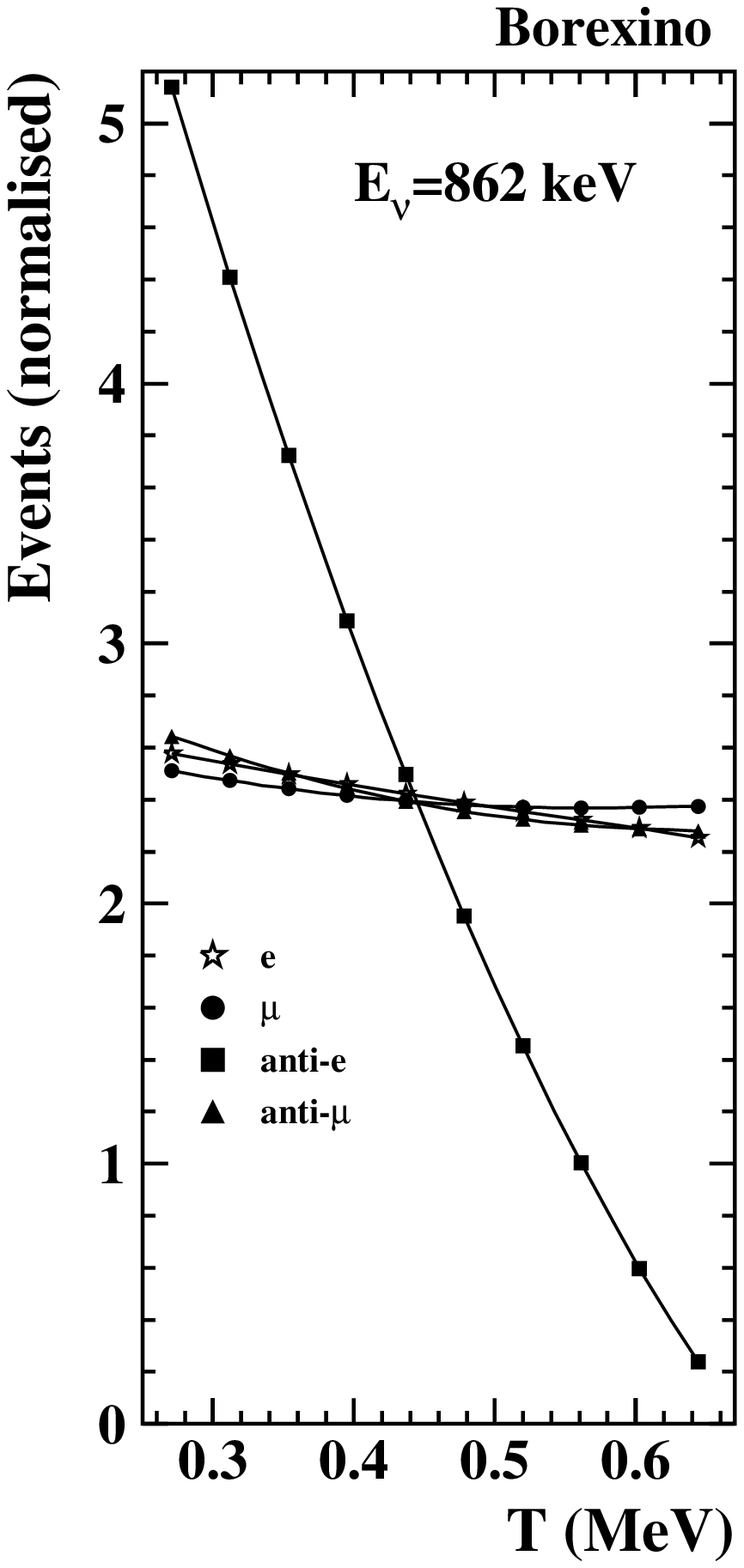,width=0.35\textwidth}
}
\caption{Normalised recoil electron kinetic energy distributions, 
for each of the 4 neutrino energy bins (see text) at HELLAZ (left) 
and for $^7$Be neutrinos (right).}
\label{distributions}
\end{figure}

The issue we would like to concentrate on is whether the shapes of the
recoil electron kinetic energy distributions for different (anti)neutrino
species are statistically different at Borexino and HELLAZ. With this in
mind, Fig.~\ref{distributions} depicts the normalised distribution of events
at HELLAZ (left) and Borexino (right).\footnote{In addition to $pp$-neutrinos,
HELLAZ is also sensitive to $^7$Be neutrinos, as well as the 
$pep$-neutrinos and
the neutrinos coming from the CNO-cycle. $^7$Be neutrinos can be clearly 
separated from $pp$-neutrinos, while the number of $pep$ and CNO-cycle
neutrino generated events is expected to be less than 10\% that of 
$pp$-neutrinos. Borexino is sensitive to, in addition to $^7$Be neutrinos 
(with $E_{\nu}=862$~keV), a fraction of the $pep$ and the CNO-cycle neutrinos,
which produce approximately 10\% as many events as $^7$Be neutrinos. 
We assume throughout, for simplicity, that only $pp$ ($^7$Be) neutrinos 
are detected at HELLAZ (Borexino).}
In the case of Borexino, the data
is binned into ten kinetic energy bins, between 250~keV and 650~keV. In the
case of HELLAZ, the data is binned into $4\times 21$ bins in $E_{\nu}\times T$.
The bins have a width of 50~keV in the $E_{\nu}$ direction 
and central values of 245, 295, 345 and 395~keV, while in the $T$ direction
they have a width of 10~keV in the range from 50 to 260~keV. 
The bin sizes have been chosen such that they are roughly the same as 
the resolution of both detectors. In order to 
integrate over the incoming neutrino energy at HELLAZ, the (normalised) BP98  
$pp$-spectrum presented at \cite{bahcall_www} was used.

Many important features of the recoil electron kinetic distributions are
worthwhile to point out. First of all, it is quite clear that the spectrum 
produced by $\bar{\nu}_e$-$e$ is much steeper than all the other 
ones.\footnote{Some of these features were pointed out in \cite{anti_nus}.}
Second, the $\nu_e$ and $\nu_{\mu}$ generated spectra have opposite 
slopes when the neutrino energy is small enough, while their shapes start to 
look more and more similar as the neutrino energy increases. Finally, the 
spectra produced by $\nu_{\mu}$-$e$ and $\bar{\nu}_{\mu}$-$e$ scattering are 
extremely similar, especially at very low energies.

All of these features can be readily understood from Eq.~(\ref{eq_dist}). First,
it is convenient to write the expression for the normalised recoil electron
kinetic energy distributions, 
\begin{equation}
\frac{{\rm d}\bar{\sigma}}{{\rm d}y}=N\left[\frac{A}{B}+\frac{B}{A}
\left( 
1-y\right)^2-\frac{m_e}{E_{\nu}}y, 
\right],
\label{eq_dist_norm}
\end{equation} 
where $y=T/E_{\nu}$, and $N$ is a normalisation constant, such that 
$\int\frac{{\rm d}\bar{\sigma}}{{\rm d}T}$ from $T_{\rm min}$ to $T_{\rm max}$ 
equals unity. 

For $\nu_{e}$, $\frac{A}{B}\sim 3$, while $\frac{B}{A}\sim 1/3$. 
For $\bar{\nu}_e$ the situation is reversed, while for
both $\nu_{\mu}$ and $\bar{\nu}_{\mu}$,
$\frac{A}{B}\sim\frac{B}{A}\sim -1$.\footnote{In this case, the
  normalisation constant $N$ is negative.} 
Note, curiously enough, that the reason for the similarity between the 
$\nu_{\mu}$ and $\bar{\nu}_{\mu}$ cases is simply due to the accidental fact
that $\sin^2\theta_W$ is close to $1/4$.\footnote{The fact that there is 
a sign difference between $g_L$ and $g_R$ for muon-type (anti)neutrinos is
irrelevant, since these coefficients either appear as squares or as the product
$g_Lg_R$.} This similarity is even more pronounced at very low energies,
when the $\frac{m_e}{E_{\nu}}$ term dominates over the $\frac{A}{B}$ and 
$\frac{B}{A}$ terms. 
    
Keeping in mind that $0.59\lesssim\frac{m_e}{E_{\nu}}\lesssim 2.3$ in
the neutrino 
energy range of interest and $0<y\leq (1+m_e/2E_\nu)^{-1} <1$, 
one may write approximate expressions
\begin{eqnarray} 
&\left(\frac{{\rm d}\bar{\sigma}}{{\rm d}y}\right)_{\nu_e}\propto
3-\frac{m_e}{E_{\nu}}y, \\
&\left(\frac{{\rm d}\bar{\sigma}}{{\rm d}y}\right)_{\bar{\nu}_e}\propto
3(1-y)^2-\frac{m_e}{E_{\nu}}y, \\
&\left(\frac{{\rm d}\bar{\sigma}}{{\rm d}y}\right)_{\bar{\nu}_{\mu}}\propto
\left(\frac{{\rm d}\bar{\sigma}}{{\rm d}y}\right)_{\nu_{\mu}}\propto
1+(1-y)^2+\frac{m_e}{E_{\nu}}y.
\end{eqnarray}  
In the limit $\frac{m_e}{E_{\nu}}\ll 1$, all three distributions are quite
different (see, {\it e.g.,}\/ Fig.~1(B) in \cite{recoil_spectrum}). The $\nu_e$
case is roughly flat, the $\bar{\nu}_e$ case ranges from 3 at $y=0$ to 0 at 
$y=1$ and the $\nu_{\mu},\bar{\nu}_{\mu}$ case ranges from $3/2$ at $y=0$ to 
$3/4$ at $y=1$. 

For $\frac{m_e}{E_{\nu}}\gtrsim 1$, things are slightly more complicated, 
but still
easy to understand. For example, the slope of the distributions for small
values of $y$ are, up to normalisation factors,  
$-\frac{m_e}{E_{\nu}}$, $-\left(6+\frac{m_e}{E_{\nu}}\right)$
and $+\left(\frac{m_e}{E_{\nu}}-2\right)$ for $\nu_e$, $\bar{\nu}_e$ and 
$\nu_{\mu},\bar{\nu}_{\mu}$ respectively. 
It is then
easy to note that the $\bar{\nu}_e$ slope is significantly
more negative than the 
other two, and that, in the case of $\nu_{\mu},\bar{\nu}_{\mu}$ the slope is
actually positive if $E_{\nu}$ is small enough. This is indeed what one
observes in Fig.~\ref{distributions}. 

As the incoming neutrino energy increases, the distributions generated by 
$\nu_e$ and $\nu_{\mu},\bar{\nu}_{\mu}$ look more and more similar. One hint
of this behaviour is that the slope of the $\nu_e$ case increases (decreases in
absolute value), while the slope of the    
$\nu_{\mu},\bar{\nu}_{\mu}$ decreases. One can easily estimate that for 
$0.9$~MeV$\lesssim E_{\nu}\lesssim 1.0$~MeV the shapes of the $\nu_e$ and 
$\nu_{\mu}$ induced recoil kinetic energy distributions are most similar. 
Indeed, for $^7$Be neutrino energies, one can already note that the difference
between the $\nu_e$ and the $\nu_{\mu}$ cases is similar to the difference
between the $\nu_{\mu}$ and the $\bar{\nu}_{\mu}$ cases.
 
   
\setcounter{footnote}{0}
\section{Measuring a $\nu_{\mu,\tau}$ Component in the Solar Neutrino
  Flux\label{sec:measuring}} 

In this section, we address the question whether the shapes of the 
recoil electron kinetic energy distributions presented in Sec.~2 are 
{\sl statistically different}\/ 
at Borexino or HELLAZ. In
the affirmative case, there is hope that one may be sensitive to a  
``contamination'' of other neutrino types in the solar neutrino flux by
analysing the shape of the recoil kinetic energy spectrum.  We
consider this an ``appearance experiment'' of the ``wrong'' types
of neutrinos from the Sun.  In this section we will only consider
the case of $\nu_e\leftrightarrow\nu_{\mu}$ oscillations.  
 
In the advent of neutrino oscillations, a mixture of different neutrino
weak eigenstates reaches the Earth. 
Given an electron-type neutrino survival probability 
$P_{ee}$, a fraction $P_{ee}$ of all the neutrinos arriving at the
detector are $\nu_e$, while a fraction $1-P_{ee}$ are $\nu_{\mu}$. The 
recoil electron kinetic energy distribution will, therefore, be given by
\begin{equation}
\frac{{\rm d}\sigma(T,E_{\nu})}{{\rm d}T}=P_{ee}\times\left(
\frac{{\rm d}\sigma(T,E_{\nu})}{{\rm d}T}\right)_{\nu_e}+(1-P_{ee})\times\left(
\frac{{\rm d}\sigma(T,E_{\nu})}{{\rm d}T}\right)_{\nu_{\mu}}.
\label{eq:data}
\end{equation}
Note that, in general, $P_{ee}$ is a function of the neutrino 
oscillation parameters (the mass-squared differences of the neutrino 
mass eigenstates and the neutrino mixing angles) and the neutrino energy.

We simulate ``data'' at Borexino and HELLAZ for different values of $P_{ee}$. 
We use the distributions presented in Sec.~2, while the flux
of $pp$ and $^7$Be neutrinos are taken from the SSM \cite{SSM}. In the case of
Borexino, the energy dependence of $P_{ee}$ is irrelevant, given the monochromatic
nature of $^7$Be neutrinos. In the case of HELLAZ, we assume that $P_{ee}$ is
constant inside each individual neutrino energy bin. 
Following the central idea presented in \cite{recoil_spectrum}, we perform a 
$\chi^2$
fit to the ``data'' using a linear combination of $\nu_e$-$e$ scattering
and $\nu_{\mu}$-$e$ scattering with arbitrary coefficients,
\begin{equation}
C_{e}\times\left(
\frac{{\rm d}\sigma(T,E_{\nu})}{{\rm d}T}\right)_{\nu_e}+C_{\mu}\times\left(
\frac{{\rm d}\sigma(T,E_{\nu})}{{\rm d}T}\right)_{\nu_{\mu}},
\label{eq:trial}
\end{equation}
{\it i.e.,}\/ we
perform a two parameter ($C_e$ and $C_{\mu}$)
fit to the data. This measurement procedure is
independent of the SSM prediction for the neutrino flux. Therefore, if a
nonzero coefficient of the $\nu_{\mu}$-$e$ scattering distribution is measured,
one can claim to have detected evidence for neutrinos other than $\nu_e$ 
coming from the Sun. This ``appearance'' result certainly qualifies as a smoking
gun signature for neutrino oscillations.       

\begin{figure} [h]
\centerline{
  \psfig{file=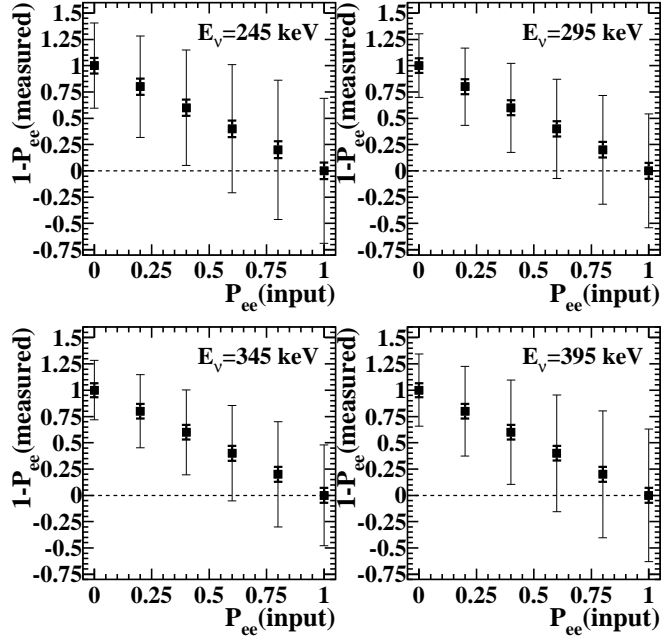,width=0.7\textwidth}
}
\caption{The ``measured'' value of $1-P_{ee}$ as a function of the input value
of $P_{ee}$ for each of the 4 neutrino energy bins (see text), after 5 years of
HELLAZ running. The long, thin error bars correspond to
model-independent analyses based only on the electron recoil energy spectrum
shape, while the short, thick ones correspond to analyses 
which include the SSM prediction for the solar $pp$-neutrino flux, with
an (inflated) uncertainty of 5\%.}
\label{measurement_1}
\end{figure}

Fig.~\ref{measurement_1} (long, thin error bars) depicts the measured 
value of $1-P_{ee}=\frac{C_{\mu}}{C_e+C_{\mu}}$ 
in each of the neutrino energy bins defined in Sec.~2 as
a function of the input value of $P_{ee}$, for 5 years of 
simulated HELLAZ data. As was mentioned in the previous paragraph, 
the relevant information one should obtain from the plot is if the measured 
value of $1-P_{ee}\propto C_{\mu}$ is 
statistically different from zero.   

\begin{figure} [h]
\centerline{
  \psfig{file=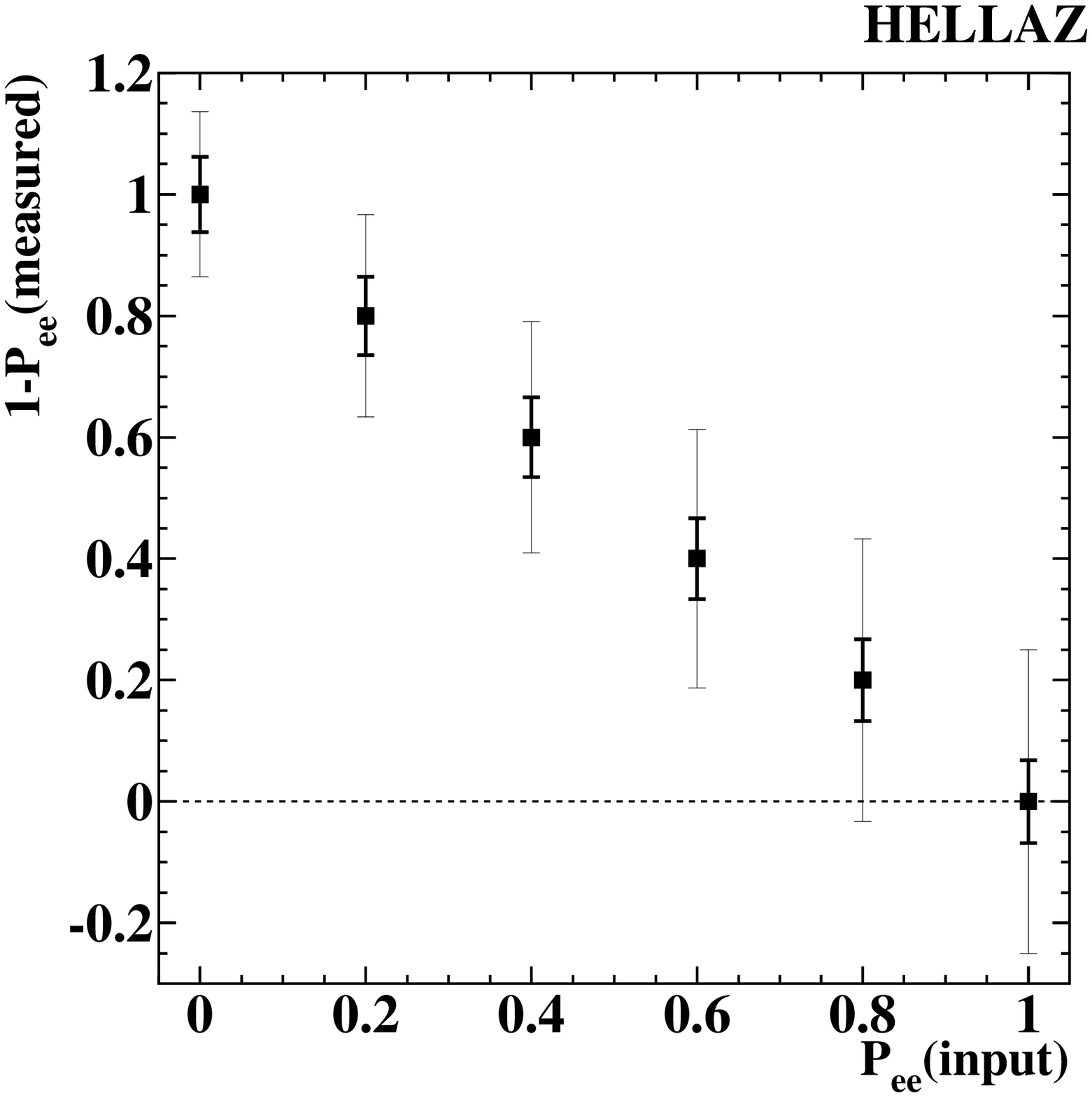,width=0.5\textwidth}
  \psfig{file=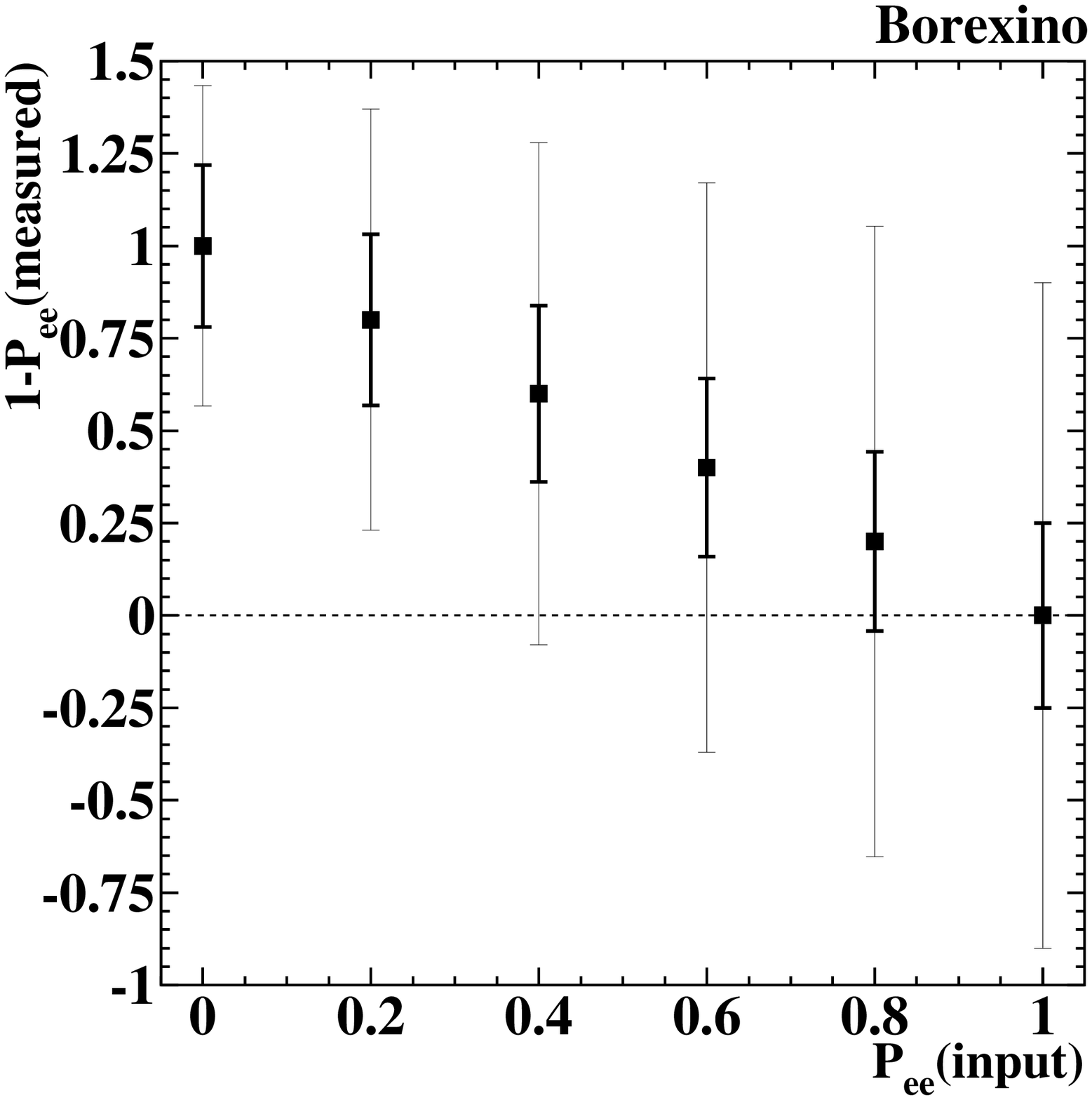,width=0.5\textwidth}
}
\caption{The ``measured'' value of $1-P_{ee}$ as a function of the input value
of $P_{ee}$, after 5 years of HELLAZ (left) and 2 years of Borexino running
(right). $P_{ee}$ is assumed to be constant for 
$E_\nu = $220--420~keV in the case of HELLAZ.  
The long, thin error bars correspond to
model-independent analyses based only on the electron recoil energy spectrum
shape, while the short, thick ones correspond to analyses
which include the SSM prediction for the solar $pp$ ($^7$Be) neutrino flux, with
an (inflated) uncertainty of 5\% (20\%).}
\label{measurement_2}
\end{figure}

Fig.~\ref{measurement_2} (right, long, thin error bars) 
depicts the measured value of $1-P_{ee}$ as
a function of the input value of $P_{ee}$, for two years of Borexino
running. This is just a repetition of Fig.~2(A) in 
\cite{recoil_spectrum}.\footnote{In 
\cite{recoil_spectrum}, a different variable, $P\equiv 1-P_{ee}$, 
was used. Both results are, of course, equivalent.} 
Fig.~\ref{measurement_2} (left, long, thin error bars) 
depicts the result obtained at HELLAZ if all
energy bins are used in the ``data'' analysis. This result is only meaningful 
if $P_{ee}$ is roughly constant for neutrino energies ranging from from
220~keV to 420~keV. 
This happens to be the case for most of the currently preferred
regions of the two-neutrino oscillation parameter space, 
especially LMA, LOW and VAC solutions (see, {\it e.g.}\/, 
\cite{bksreview}).\footnote{For the SMA solution, there is a sharp 
drop in $P_{ee}$ at $E_{\nu} \simeq 0.4$~MeV, and the three lower bins 
can be combined without any problem.  At HELLAZ, this will show up in 
the data, as the $E_{\nu}$ spectrum differs from the expected $pp$-neutrino 
spectrum, and hence is not a concern.}
Clearly, the significance of the measurement is better 
than the one obtained for individual energy bins (Fig.~\ref{measurement_1}).   

Next, the same analysis as above is repeated, except that the SSM prediction
for the solar neutrino flux is included in the $\chi^2$ analysis. An uncertainty 
of 20\% (5\%) was assumed for the $^7$Be ($pp$) neutrino flux. The theoretical
error was considered Gaussian for simplicity.\footnote{This procedure follows
the one used in \cite{seasonal}. The readers are referred to this article for 
details.} Note that the uncertainties assumed here are inflated with respect to
the ones quoted in the SSM calculations \cite{SSM} (9\% and 1\%, respectively), 
in order to render the results very conservative.  Since
the rates are very high both at Borexino and HELLAZ, the error bars
are dominated by the uncertainties in the fluxes and hence they can be
approximately scaled according to the assigned flux
uncertainties.

The results are presented in Figs.~\ref{measurement_1} and \ref{measurement_2}
(short, thick error bars). The significance of the measured value of 
$1-P_{ee}$ improves
significantly, especially at HELLAZ, because of the small assigned uncertainty on the 
$pp$-neutrino flux. After five years of HELLAZ running, for example, 
one should be be able to determine a 1-sigma-away-from-zero $\nu_{\mu}$ 
component in the $pp$-neutrino flux even for $P_{ee}\sim0.9$. It is also 
noteworthy that in the case of the SMA MSW solution to the solar 
neutrino puzzle $P_{ee}\sim0$ for $^7$Be neutrinos,
in which case a 4-sigma-away-from-zero evidence for $\nu_{\mu}$ in the solar
neutrino flux can be established in only two years of Borexino running!    
It is important to emphasise that using SSM predictions for the solar neutrino
flux is a reasonable thing to do, especially for $pp$-neutrinos. As mentioned 
before, the flux of $pp$-neutrinos is very well known because it is tightly
related to the flux of light coming from the Sun. It is, therefore, the neutrino
flux which is least sensitive to detailed modelling of the Sun's innards.

Some comments are in order. First, only statistical uncertainties were 
considered,
and there are no background events in our ``data.'' 
As discussed in Sec.~\ref{sec:recoil}, 
the assumption of a 
negligible background rate seems less than realistic at Borexino, but
may be possible in future upgrades. It may, however, be a 
fair assumption in the case of HELLAZ.
If the real experimental data contains
a sizable number of background events, it is necessary to either subtract
the background in a bin-by-bin basis or to somehow model the recoil kinetic energy 
distribution produced by background events. Analysing either of these procedures, 
however, is beyond the scope of this paper.   
  
Second, the analysis which does not include the SSM flux predictions
is completely model-independent (the only assumption being the
electron recoil spectrum as predicted by the standard electroweak
theory), while the one which includes the SSM flux predictions is
model-dependent.  Obviously, one obtains a much better determination
of $P_{ee}$ with the additional input of the SSM flux predictions.  
For establishing the
``wrong neutrino component'' in the solar neutrino flux as a smoking
gun signature of the solar neutrino oscillations, the former
approach is desired.  However, for the purpose of determining the
oscillation parameters, the energy dependence of the survival
probability, and excluding other neutrino oscillation modes, such as 
$\nu_e\leftrightarrow\nu_s$ or $\nu_e\leftrightarrow\bar{\nu}_{e,\mu,\tau}$
(as will be discussed in Sec.~\ref{sec:testing}), it
is reasonable to include the SSM predictions in the analysis.  

Finally, we point out that the results we obtained for HELLAZ are similar to
the ones obtained by J.~S\'{e}guinot {\it et al}\/ \cite{spectrum_first}. Indeed, 
we chose neutrino energy bins at HELLAZ which coincide with the ones used in 
\cite{spectrum_first}. They also perform two different analyses of their simulated
data, one which is independent of the SSM prediction for the solar neutrino flux,
and one which assumes the SSM prediction for the flux. However,
their data analysis procedure is somewhat different, and they do not take 
the theoretical uncertainty of the solar neutrino flux prediction into account.   

\setcounter{footnote}{0}
\section{Testing for the $\nu_e\leftrightarrow\nu_s$
  or $\bar{\nu}_{e,\mu,\tau}$ Hypotheses\label{sec:testing}}

Although it is most natural to assume that electron-type neutrinos oscillate into
some linear combination of muon-type and tau-type neutrinos, there is a
logical possibility that electron-type neutrinos might oscillate into 
standard model singlet sterile neutrinos \cite{sterile_hyp}, or, perhaps,
into antineutrinos of all flavours\footnote{The original neutrino oscillation
paper by Bruno Pontecorvo \cite{Pontecorvo} did, after all, consider
 $\nu_{e}\leftrightarrow\bar{\nu}_e$!} (see \cite{anti_nus} and
references therein). In this section, we will address the issue of excluding
these solar neutrino oscillation modes {\sl if the data collected at Borexino
and HELLAZ are consistent with $\nu_e\leftrightarrow\nu_{\mu}$ oscillations}. 

One can already address these ``exotic'' oscillation modes
with the current experimental data. The flux of electron-type anti-neutrinos from 
the Sun is particularly constrained by the SuperKamiokande and the LSD experiments 
\cite{constraint_anti}: 
the 95\% CL SuperKamiokande upper bound on the flux of $\bar{\nu}_e$ from
the Sun with energies $\gtrsim 6.5$~MeV is 
$\Phi_{\bar{\nu}_e}< 1.8\times 10^5$~cm$^{-2}$
s$^{-1}$, or $\Phi_{\bar{\nu}_e}/\Phi^{^8{\rm B}}_{SSM}=0.035$, 
where $\Phi^{^8{\rm B}}_{SSM}$ is the SSM
prediction for the $^8$B neutrino flux.  KamLAND, being a dedicated
detector for $\bar{\nu}_e$, will improve this 
bound further down to 0.1\% of the SSM flux above reactor
anti-neutrino energies ($E_\nu \gtrsim 8$~MeV) after one year of running
\cite{KamLAND}. 
There are, however, scenarios in which,
for energies below the SuperKamiokande threshold, the $\nu_e\leftrightarrow
\bar{\nu}_e$ mixing is quite large (\cite{anti_nus} and references therein).
Such a possibility can only be addressed by low-energy solar neutrino experiments.  

Below, we discuss the exclusion of electron-type neutrino oscillations 
into sterile neutrinos or into one of the antineutrino types separately at
Borexino and HELLAZ experiments. 

\subsection{$\nu_e \leftrightarrow \nu_s$}

The $\nu_e\leftrightarrow\nu_s$ oscillation mode is allowed by the analysis
of current solar neutrino data \cite{rate_analysis}, even though in the case of
the MSW solutions to the solar neutrino puzzle, only the equivalent of the SMA
MSW solution exists at the 99\% confidence level \cite{rate_analysis}.  
It is curious to note that, in the case of atmospheric 
neutrinos, the $\nu_{\mu}\leftrightarrow\nu_s$ hypothesis is currently 
somewhat disfavoured \cite{atmos_sterile}.

In the case of $\nu_e\leftrightarrow\nu_s$ oscillations, one expects the
recoil electron kinetic energy spectrum to be exactly the same as the one
generated by $\nu_e$-$e$ scattering, since $\nu_s$ do not interact with
electrons. The only effect of the neutrino oscillations would be to suppress
the expected number of events, {\it i.e.}\/, the hypothesis of 
$\nu_e\leftrightarrow\nu_s$ oscillations is identical to assuming that the
solar neutrino flux is, somehow, suppressed. Therefore, we attempt to fit
the ``data'' simulated according to Eq.~(\ref{eq:data}) in Sec.~3 (remember that
the ``data'' is consistent with $\nu_e\leftrightarrow\nu_{\mu}$ oscillations) 
to the trial function Eq.~(\ref{eq:trial}), where the piece which 
corresponds to $C_{\mu}$ vanishes identically.  
This is a one parameter $\chi^2$ fit to $C_{e}$.
Note that the only discrimination against $\nu_{s}$ is the recoil 
energy spectrum, because the rate can be always fitted with the free 
parameter $C_{e}$.  The inclusion of the SSM flux prediction does not
help in excluding the $\nu_e \rightarrow \nu_s$ oscillation because
the free parameter $C_e$ makes the predicted flux
irrelevant.\footnote{One can ``discover'' $\nu_s$
  by observing a nearly vanishing rate for $^7$Be neutrinos. 
  In the case of $\nu_{e}\leftrightarrow \nu_{\mu,\tau}$ oscillations,
  neutral-current scattering guarantees at least 
  (when $P_{ee}=0$) 21\% of the SSM rate.  
  For this purpose, one should rely on the SSM flux prediction.} 
Note that
$^7$Be 
neutrinos are predicted to have almost completely oscillated to $\nu_s$
for the (only available) SMA solution: $P_{ee} =
0.009^{+0.244}_{-0.005}$ \cite{bksreview}.

\begin{figure} [h]
\centerline{
  \psfig{file=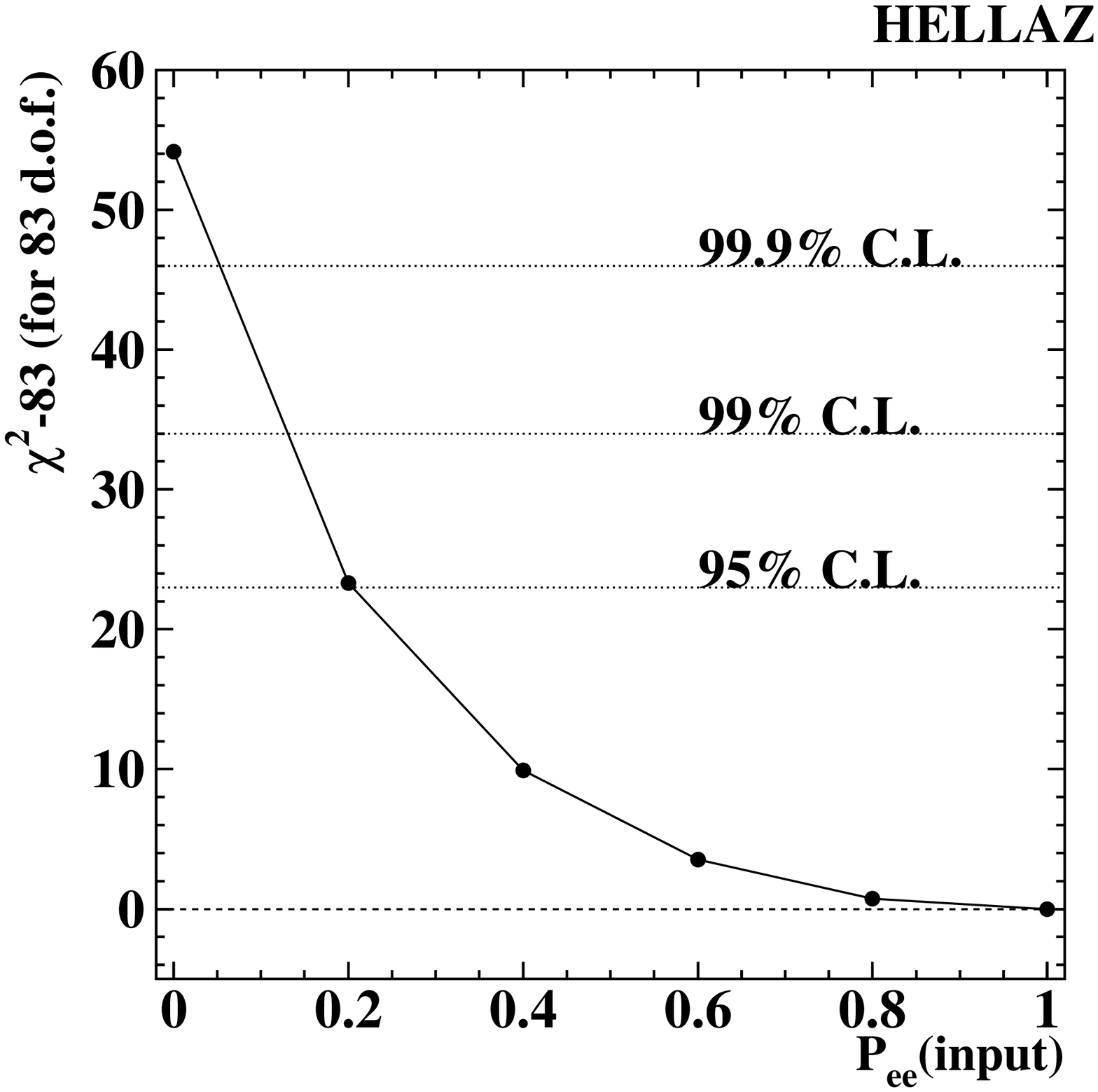,width=0.5\textwidth}
  \psfig{file=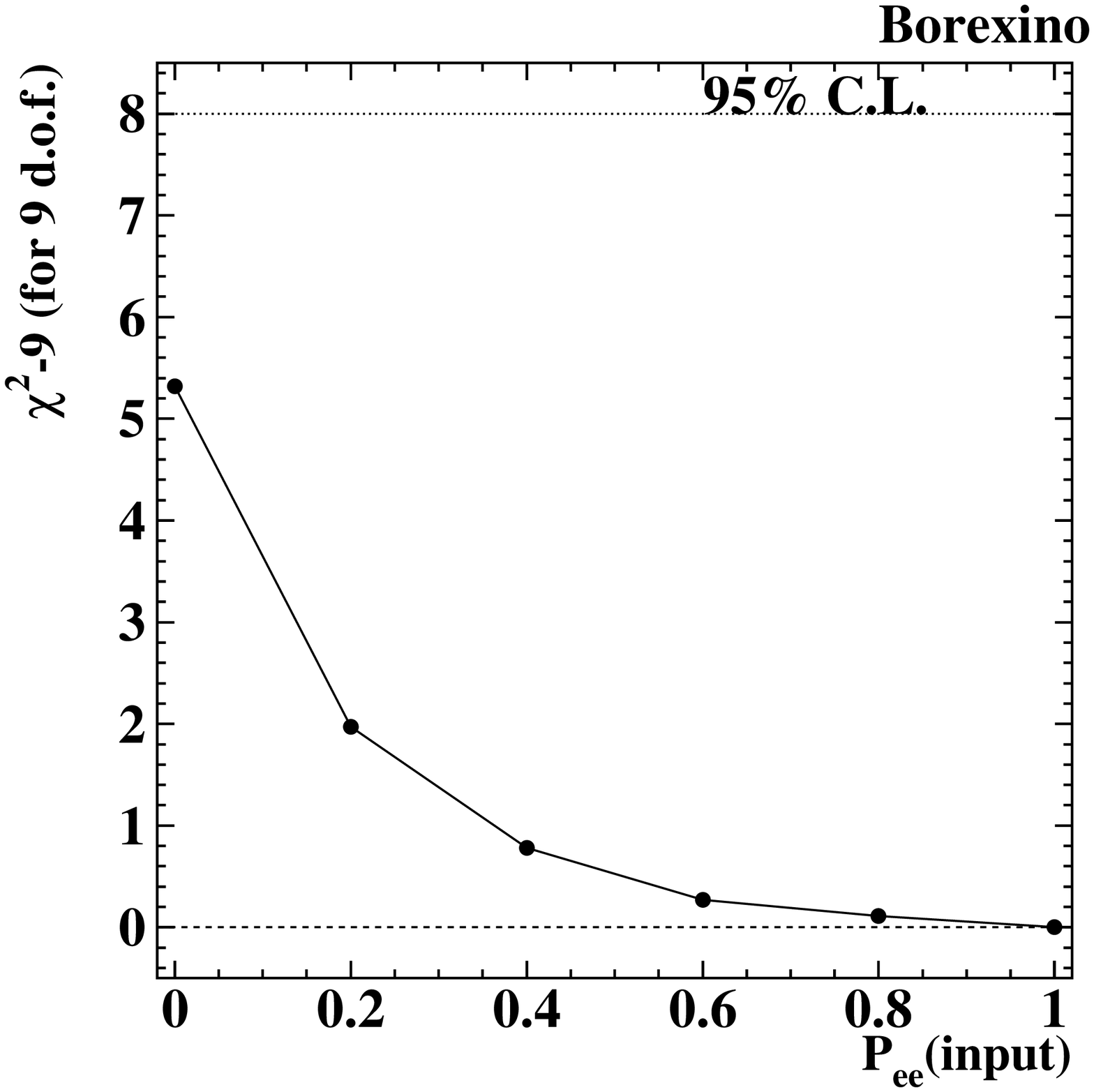,width=0.5\textwidth}
}
\caption{Minimum $\chi^2$ values as a function of the input value of $P_{ee}$, 
obtained when one tries to fit the ``data'' (which is consistent with  $\nu_e
\leftrightarrow \nu_\mu$ oscillations and $P_{ee}=P_{ee}(\rm input)$) 
with a $\nu_e + \nu_{s}$ 
distribution (see text), for 5 years of HELLAZ (left) and 2 years of 
Borexino running (right). 
The dotted lines indicate the 95\%, 99\% and 99.9\% exclusion confidence 
levels.}
\label{sterile}
\end{figure}

Fig.~\ref{sterile} depicts the value of $\chi^2$ obtained when one attempts to
fit the ``data'' to the sterile neutrino hypothesis, for 5 years of HELLAZ
running (left) and 2 years of Borexino running (right). 
In the case of HELLAZ, we have assumed 
that $P_{ee}$ is constant over the entire $pp$-neutrino energy range.
The value of $\chi^2$ is determined using the philosophy employed in 
\cite{seasonal}, and should be
compared to $N_{\rm bins}-1$ ($N_{\rm bins}$ is the number of ``data'' bins).
After 5 years of HELLAZ running, one should be able to exclude sterile neutrinos
coming form the Sun at more than 99.9\% confidence level (CL) if all electron-type
neutrino have turned into muon-type neutrinos ($P_{ee}=0$). 
After 2 years of
Borexino 
running, the sterile neutrino hypothesis is only ruled out, at best,  
at the 89\% CL.\footnote{The situation does improve, of course, it more events are 
collected at Borexino. After 5 years of Borexino running, for example, 
one can exclude sterile neutrinos for 
$P_{ee}\lesssim0.1$ at more than 95\% CL.}  
The explanation for this is the fact that the recoil electron kinetic
energy spectra are very different when one compares the $\nu_e$-$e$ and the 
$\nu_{\mu}$-$e$ scattering cases at very low energies, {\it i.e.}\/, 
$pp$-neutrinos, and similar
at $O$(MeV) energies, {\it i.e.}\/, $^7$Be neutrinos, as discussed in Sec.~2.
The exclusion CL decreases with increasing $P_{ee}$, and sterile neutrinos are
excluded after 5 years of HELLAZ running only at the 77\% CL for $P_{ee}=0.4$. 

\subsection{$\nu_e \leftrightarrow \bar{\nu}$, Model-independent 
Fit\label{subsec:modelindependent}}

\begin{figure} [h]
\centerline{
  \psfig{file=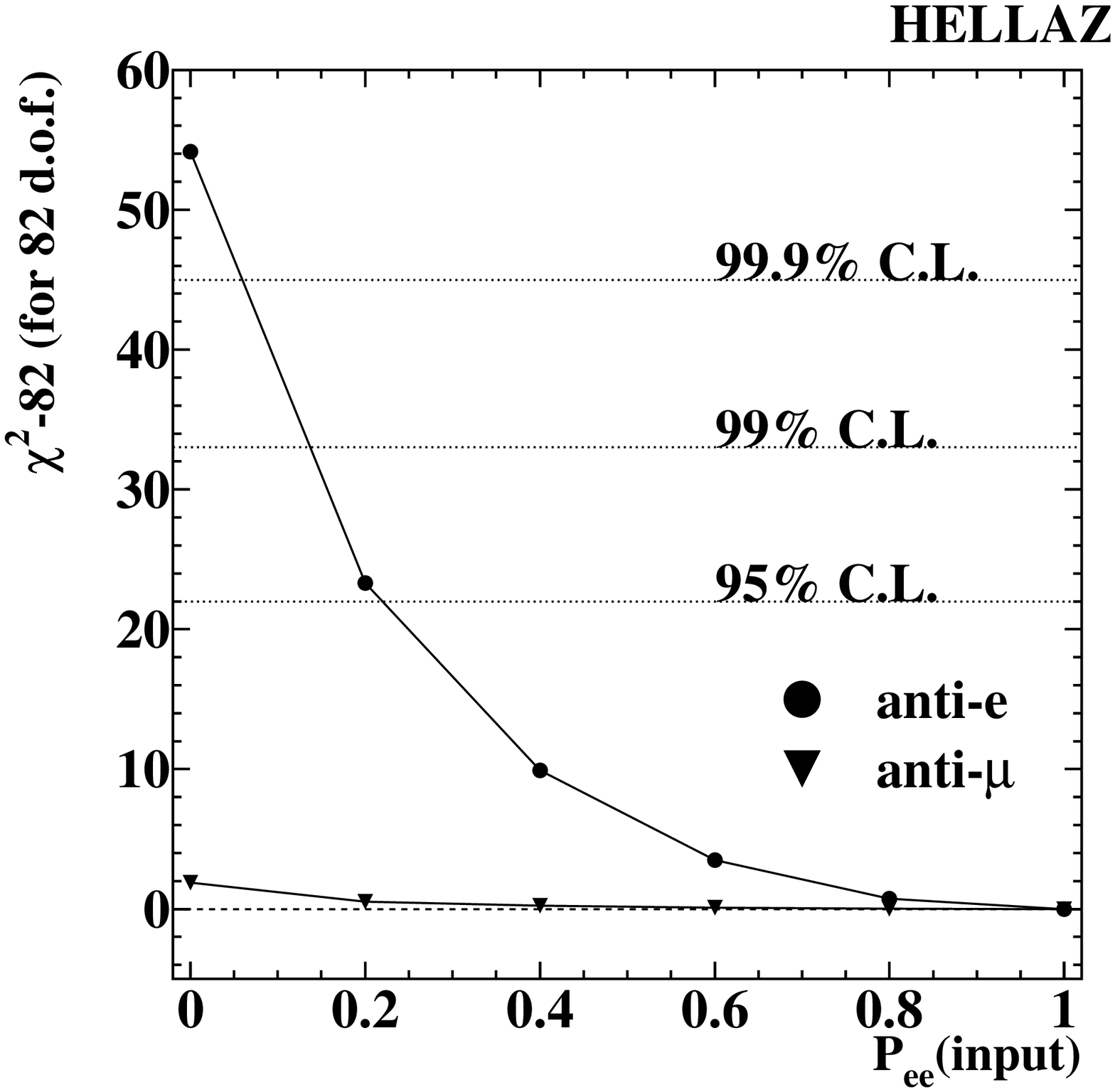,width=0.5\textwidth}
  \psfig{file=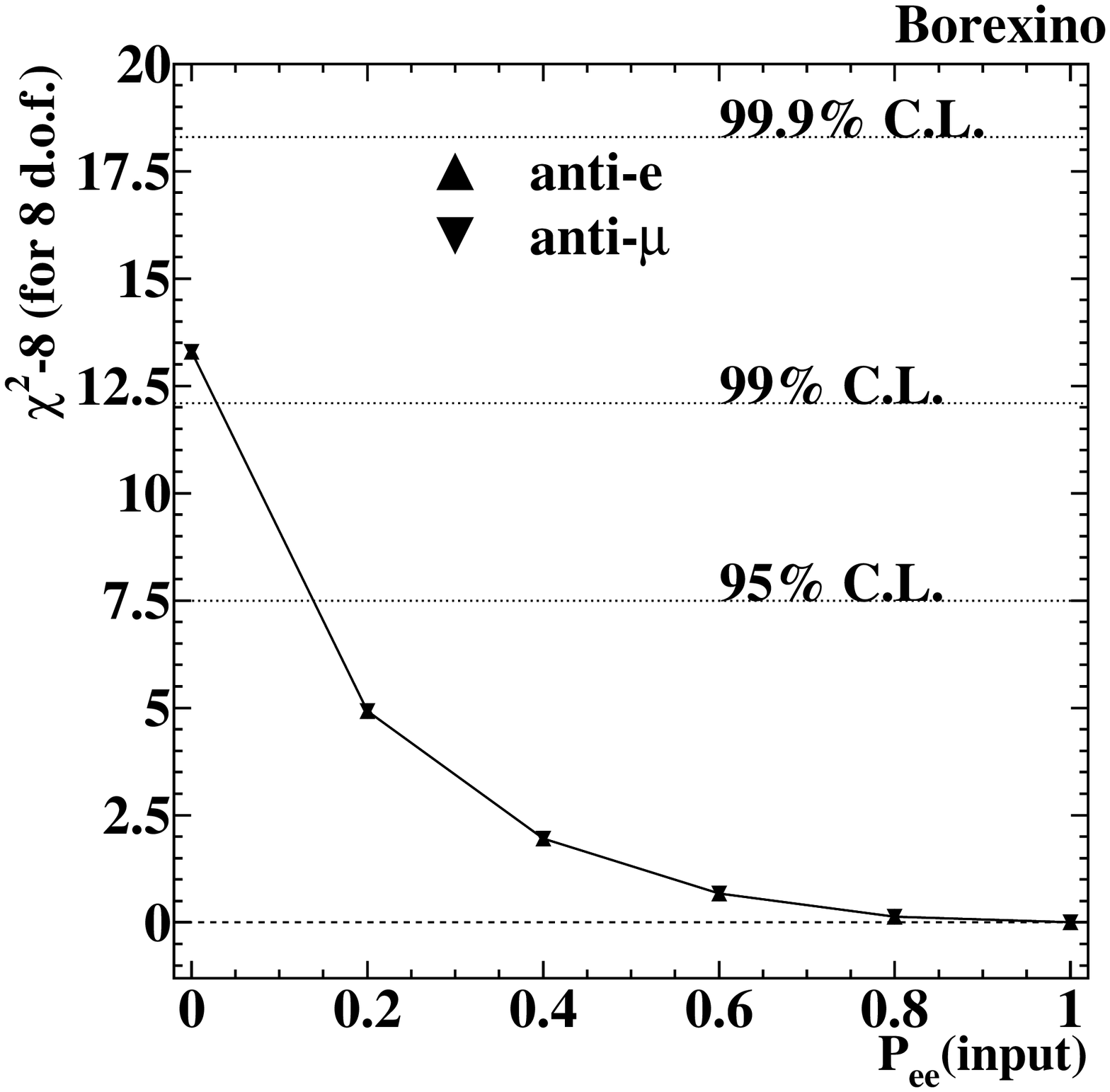,width=0.5\textwidth}
}
\caption{Minimum $\chi^2$ values as a function of the input value of $P_{ee}$, 
obtained when fitting the ``data'' with a $\nu_e + \bar\nu$ 
distribution (see text).  The fit does not include the SSM prediction for 
the solar neutrino flux and is
for 5 years of HELLAZ (left) and Borexino running (right). 
The dotted lines indicate the 95\%, 99\% and 99.9\% exclusion confidence 
levels.}
\label{anti_free}
\end{figure}

In the case of $\nu_e\leftrightarrow\bar{\nu}_{e,\mu}$ oscillations, we perform
a two parameter fit to the ``data'' simulated as in Sec.~3 to a linear combination
of the $\nu_e$-$e$ and $\bar{\nu}_{e,\mu}$-$e$ scattering recoil kinetic 
energy distributions. Fig.~\ref{anti_free} depicts the value of $\chi^2$ obtained
when such a fit is performed, for 5 years of Borexino and HELLAZ running. The value
of $\chi^2$ is to be compared to $N_{\rm bins}-2$ to determine exclusion
confidence levels. As advertised in Sec.~2, the $\nu_{\mu}$-$e$ and 
$\bar{\nu}_{\mu}$-$e$ scattering cases produce almost identical recoil kinetic
energy spectra, and are almost undistiguishable at HELLAZ. At Borexino,
however, the difference between $\nu_{\mu}$-$e$ and $\bar{\nu}_{\mu}$-$e$ 
scattering is similar to the difference between the $\nu_{\mu}$-$e$ and 
$\nu_{e}$-$e$ cases, as mentioned in Sec.~2 (see Fig.~\ref{distributions}),
and some discrimination seems possible. 
Furthermore, upon close inspection, one should note
that the shape of the distribution produced due to $\nu_e$-$e$ scattering is more
similar to the $\nu_{\mu}$-$e$ case than the $\bar{\nu}_{\mu}$-$e$. Therefore,
any $\bar{\nu}_{\mu}$ component in the trial function
makes the value of $\chi^2$ larger, {\it i.e.}\/,
the minimum of $\chi^2$ is obtained when the coefficient of the $\bar{\nu}_{\mu}$
component is zero.\footnote{We only allow nonnegative coefficients
of the distribution functions in the fits, for obvious reasons.}
This is exactly what happens in the case of $\nu_e\leftrightarrow\bar{\nu}_e$ 
oscillations, at
both experiments. Any $\bar{\nu}_e$ component in the flux makes the agreement 
between the theoretical function and the ``data'' worse, and again the best value
of $\chi^2$ is obtained when the coefficient of the $\bar{\nu}_{e}$-$e$ scattering
distribution is zero.        

One can see from Fig.~\ref{anti_free} that, after 5 years of HELLAZ data,
 $\bar{\nu}_e$ coming from the Sun can be ruled out at more than 95\% CL if 
$P_{ee}\lesssim0.2$, while $\nu_{e}\leftrightarrow\bar{\nu}_{\mu}$ oscillations 
are not constrained at all, even for $P_{ee}=0$. After 5 years of Borexino data,
both $\nu_{e}\leftrightarrow\bar{\nu}_{\mu}$ and 
$\nu_{e}\leftrightarrow\bar{\nu}_{e}$ oscillations are ruled out at more than 
95\% CL if $P_{ee}\lesssim0.1$.   

\begin{figure} [h]
\centerline{
  \psfig{file=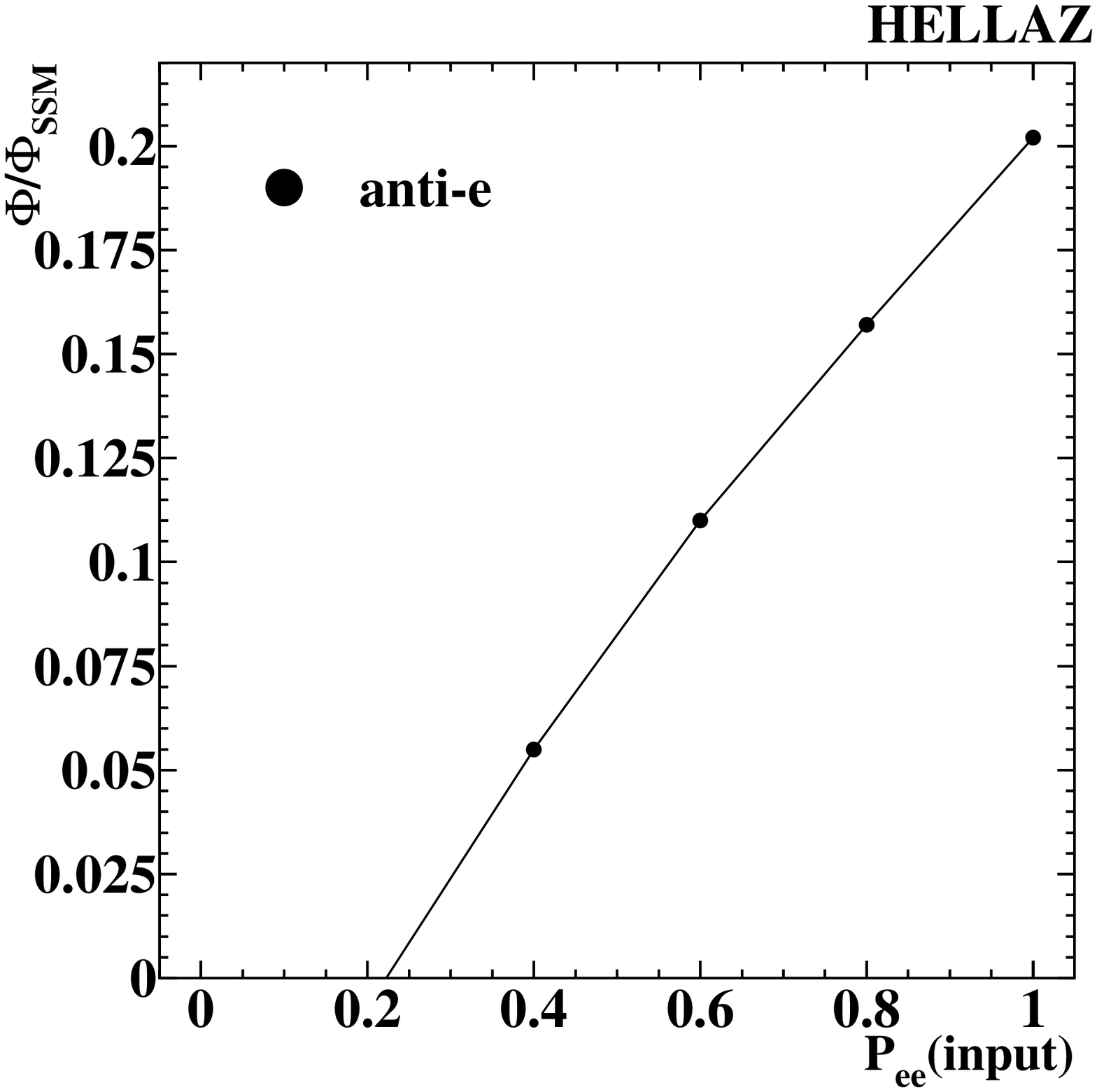,width=0.5\textwidth}
  \psfig{file=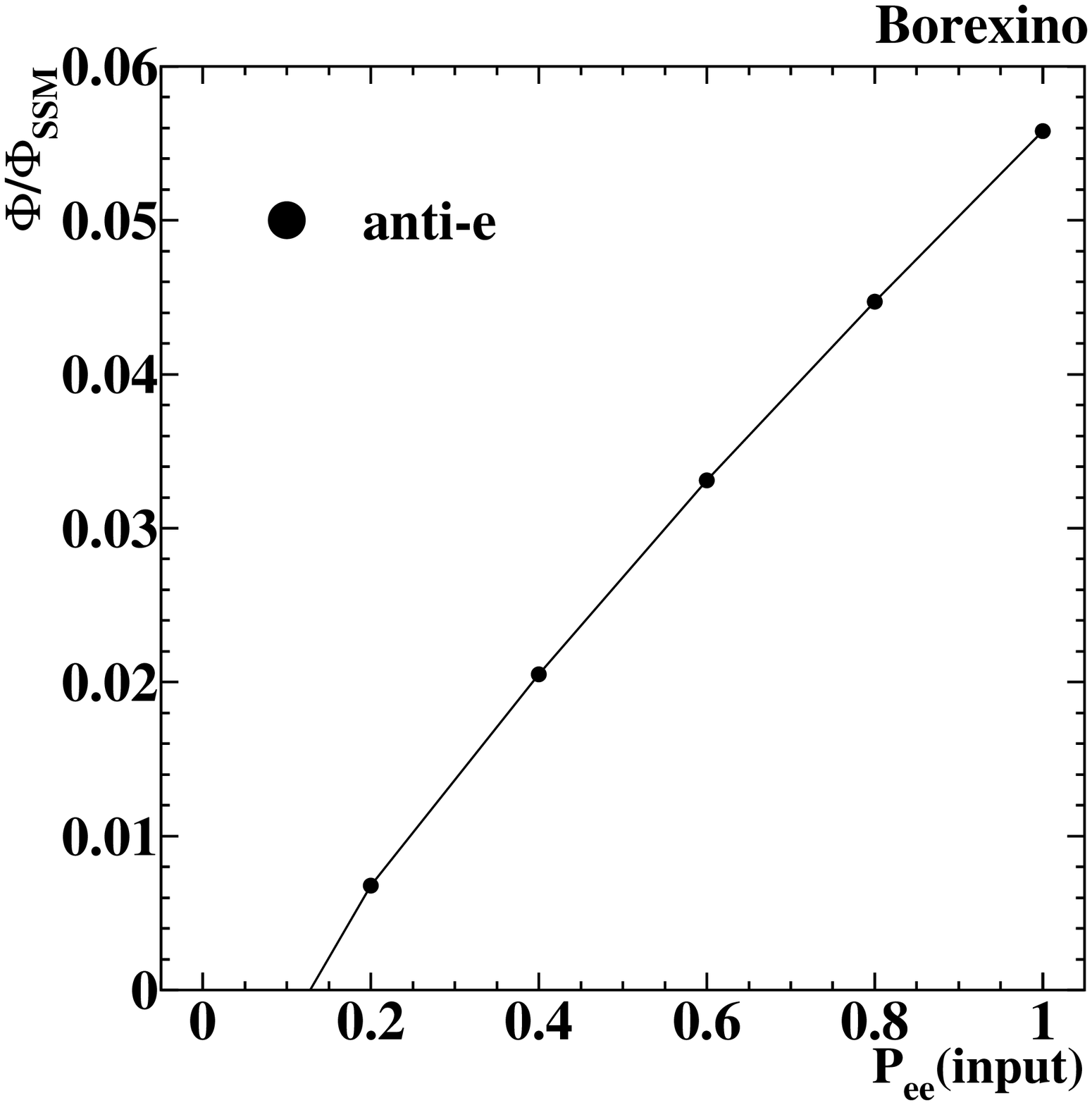,width=0.5\textwidth}
}
\caption{Upper limit on the flux of electron-type antineutrinos 
after 5 years of HELLAZ (left) and Borexino (right) running. 
The upper limits are normalised by Standard Solar Model (SSM) prediction for the 
$pp$ ($^7$Be) neutrino flux at HELLAZ (Borexino).}
\label{fluxbound}
\end{figure}

Even if the $\nu_e\leftrightarrow\bar{\nu}$ hypothesis cannot be ruled out at some
reasonable CL, one may still be able to place upper limits on the
flux of anti-neutrinos coming from the Sun. In the case of   
$\nu_e\leftrightarrow\bar{\nu}_e$ oscillations, it is straight-forward to place
 upper
bounds on the flux of electron-type antineutrinos at both HELLAZ and Borexino. The
95\% CL upper bounds on the $\bar{\nu}_e$ flux 
are depicted in Fig.~\ref{fluxbound}. Of course, for $P_{ee}\lesssim 0.2$ (0.1) at 
HELLAZ (Borexino) the upper bound on the flux is meaningless, since the {\sl 
hypothesis}\/ of $\bar{\nu}_e$ is already ruled out at more than 95\% CL. Note that
the upper bounds on the antineutrino fluxes are normalised by the SSM prediction
for the $pp$-neutrino flux $\Phi^{pp}_{SSM}=5.94\times 10^{10}$~cm$^{-2}$s$^{-1}$
for the HELLAZ result, and the SSM prediction for the $^7$Be neutrino flux 
$\Phi^{^7{\rm Be}}_{SSM}=4.8\times 10^{9}$~cm$^{-2}$s$^{-1}$ 
for the Borexino result.
For comparison, the 95\% CL
SuperKamiokande upper bound is 3.5\% of the SSM flux, while KamLAND 
will improve it to 0.1\% after one year of running.  However, both of them
are only for the $^8$B neutrinos.  Both the HELLAZ and (especially) the 
Borexino limits obtained from 5 years of ``data'' are competitive with the
SuperKamiokande limit for lower energy neutrinos.  

In the case of $\nu_e\leftrightarrow\bar{\nu}_{\mu}$ oscillations 
the situation is more 
ambiguous, especially at HELLAZ.\footnote{The same is true at Borexino if one assumes 
the SSM prediction of the total neutrino flux, as will be 
described later.} Not 
only are the minimum values of $\chi^2$ very small, but in some cases (especially
for small values of $P_{ee}$) a zero $\bar{\nu}_{\mu}$ flux is ruled out at more than
95\% CL. In such cases, it seems that the reasonable thing to do is to measure the
antineutrino flux, not determine upper limits! The only exception to this is the 
case $P_{ee}=1$, when the data looks exactly like the SSM prediction, without 
neutrino oscillations. Indeed, one can not only set upper limits on the antineutrino
fluxes, but should also set limits to the $\nu_{\mu}$ flux. Such limits are presented
in Table \ref{table}.
\begin{table}
\caption{Model-independent 95\% CL upper limits on the flux of solar 
muon-type neutrinos and antineutrinos,
when the data after 5 years of HELLAZ/Borexino running is consistent
with SSM predictions.}
\begin{center}
\begin{tabular}{|c|c|c|}
\hline
Experiment & $\Phi_{\nu_{\mu}}/\Phi_{SSM}$ & $\Phi_{\bar{\nu}_{\mu}}/\Phi_{SSM}$
 \\ \hline
HELLAZ & 1.18 & 1.44 \\ \hline
Borexino & 1.62 & 3.77 \\ \hline
\end{tabular}
\end{center}
\label{table}
\end{table}

It is worthwhile to comment that the information contained in Figs.~\ref{anti_free},
\ref{fluxbound}, and in Table~\ref{table} is also valid for the case of any 
unknown source of solar antineutrinos of the electron and the muon-types, not only 
neutrino oscillations. This is because our ``data'' was analysed assuming that the 
total flux of solar neutrinos is unknown. We emphasise that $P_{ee}$ is the 
survival probability of electron neutrinos assuming that they oscillate into active 
neutrinos, {\it i.e.,}\/ $\nu_e\leftrightarrow\nu_{\mu}$ oscillations.       

\subsection{$\nu_e \rightarrow \bar{\nu}$, SSM-dependent Fit}

\begin{figure} [h]
\centerline{
  \psfig{file=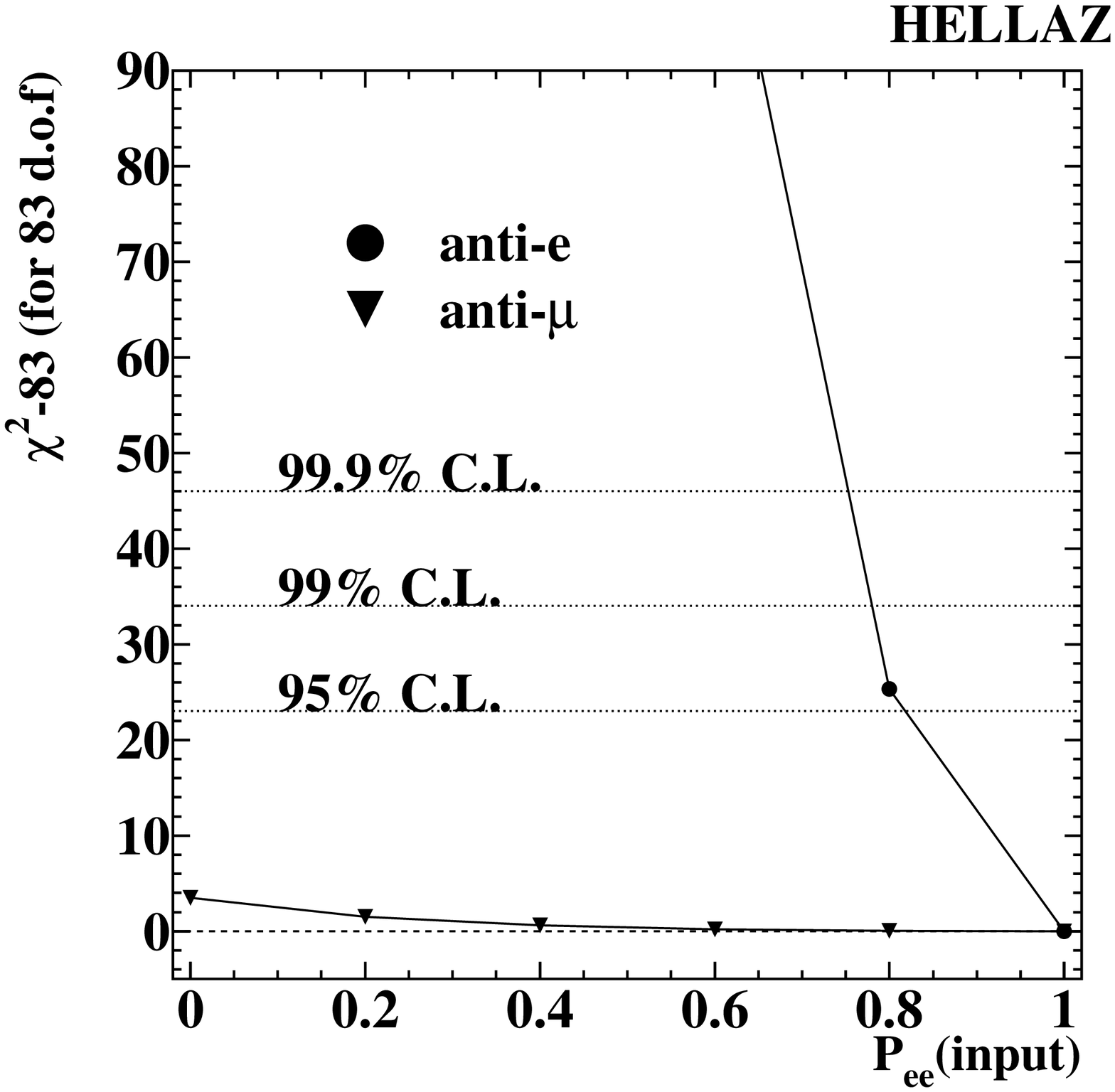,width=0.5\textwidth}
  \psfig{file=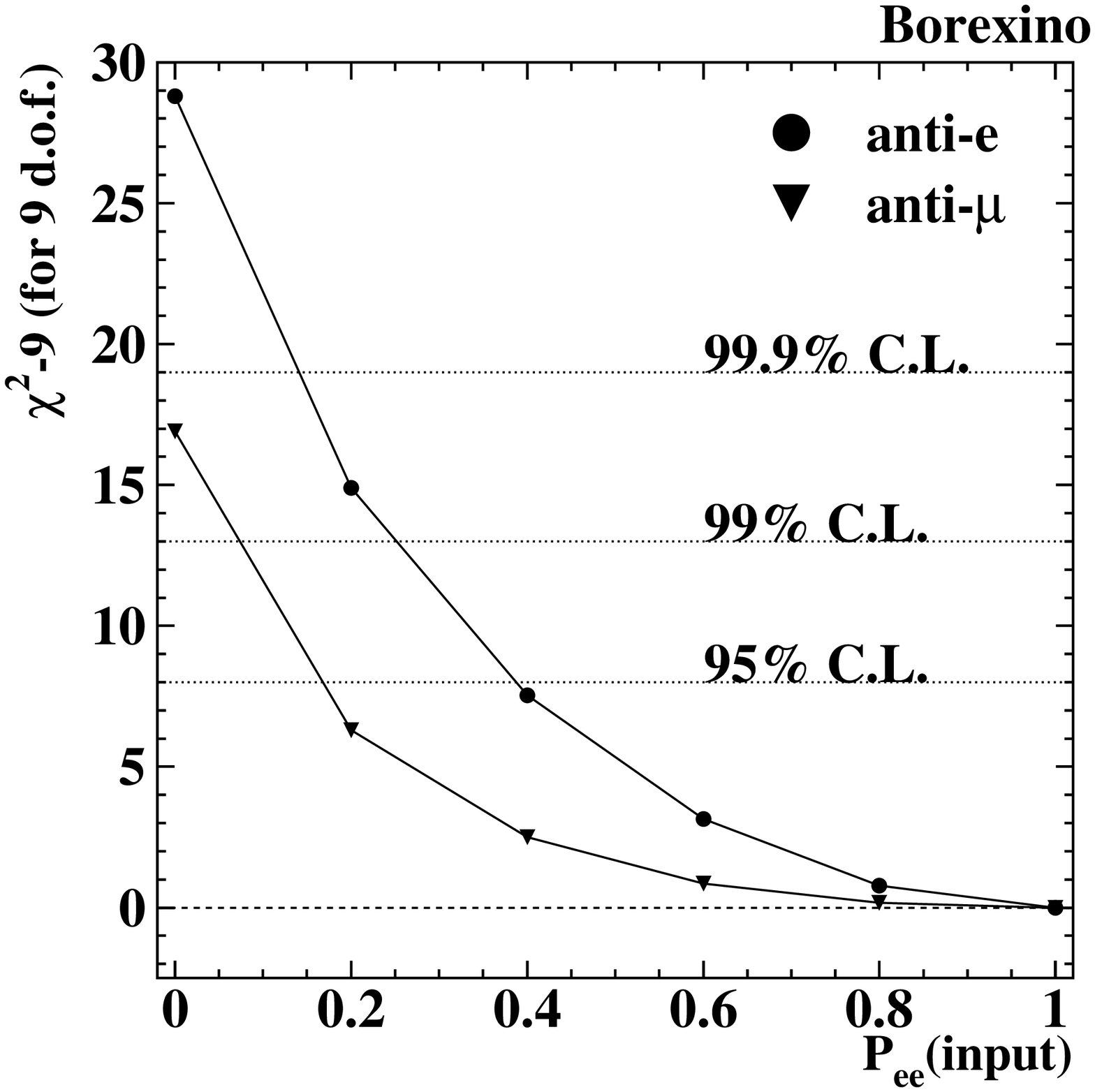,width=0.5\textwidth}
}
\caption{Minimum $\chi^2$ values as a function of the input value of $P_{ee}$, 
obtained when fitting the ``data'' with a $\nu_e + \bar\nu$ 
distribution (see text).  The fit assumes the SSM prediction for the solar neutrino
flux with an (inflated) uncertainty of 2\% (20\%) for
$pp$ ($^7$Be) neutrinos, 
for 5 years of HELLAZ (left) and Borexino running (right). 
The dotted lines indicate the 95\%, 99\% and 99.9\% exclusion confidence 
levels.}
\label{anti_flux}
\end{figure}

Next, the same analysis can be repeated assuming 
that the solar neutrino flux is known within theoretical
errors. Again, the value of $\chi^2$ is computed\footnote{This procedure follows
the one used in \cite{seasonal}. The readers are referred to this article for 
details.} and compared with $N_{\rm bins}-2+1$
(the $-2$ corresponds to the two coefficients that are varied during the minimisation 
procedure and the $+1$ corresponds to the solar neutrino flux constraint). 
Fig.~\ref{anti_flux} depicts the minimised values of $\chi^2$ obtained
with 5 years of HELLAZ (left) and Borexino (right) ``data.'' 
The theoretical uncertainty on
the $pp$ ($^7$Be) neutrino flux was taken to be 2\% (20\%); we 
inflated the theoretical errors by roughly a factor of two from those 
in BP98. 

A comparison between Figs.~\ref{anti_free} and \ref{anti_flux} reveals that the 
exclusion confidence levels increase, sometimes significantly. For example, after
5 years of HELLAZ one can exclude $\nu_e\leftrightarrow\bar{\nu}_e$ oscillations
for virtually
all values of $P_{ee}$ at more than 99.9\% CL. 
This is mostly because the $\bar{\nu}_e$ has a 
total cross section which is significantly larger than $\nu_{\mu}$, 
and the oscillation $\nu_e \leftrightarrow
\bar{\nu}_e$ cannot account for the large suppression in the event rate
in the ``data'' (due to $\nu_e \leftrightarrow \nu_\mu$).
Even the elusive 
$\nu_e\leftrightarrow\bar{\nu}_{\mu}$ case can be excluded at Borexino at more
than 95\% CL for $P_{ee}\lesssim 0.2$. 
Note that at HELLAZ the ability to discriminate
between $\nu_{\mu}$ and $\bar{\nu}_{\mu}$ is still quite limited. It is worthwhile
to comment that, unlike in the case of model-independent fits
in Sec.~\ref{subsec:modelindependent}, the minimum value of 
$\chi^2$ 
is in general obtained for a nonzero coefficient of the $\bar{\nu}$-$e$ scattering
distribution. The reason for this is that, even though the shape of the
$\bar{\nu}$-$e$ scattering recoil electron kinetic energy distribution is ``more
wrong,'' the contribution to the overall cross section is smaller than the 
$\nu_e$-$e$ scattering case, and therefore one obtains values of the solar neutrino
flux which are closer to the theoretical ones by having a finite 
$\bar{\nu}$ component, decreasing the value of $\chi^2$. 

Again, one may set upper limits on the antineutrino flux. As before, there is
some ambiguity with regard to setting upper limits for the $\bar{\nu}_{\mu}$ flux,
because for almost all values of $P_{ee}\neq 1$ at both experiments a zero flux
is excluded at more than 95\% CL. On the other hand, the $\nu_{e}\leftrightarrow
\bar{\nu}_e$ oscillation hypothesis is almost completely ruled out by HELLAZ and the
upper limits obtained at Borexino are not much better than the ones depicted in
Fig.~\ref{fluxbound}. For this reason, the equivalent of Fig.~\ref{fluxbound} in 
the case at hand is not presented.

Table~\ref{table_f} contains the obtained upper limits on the (anti)neutrino 
fluxes when $P_{ee}=1$, {\it i.e.,}\/ when the data agrees with the predictions
of the SSM. Unlike the case of a free total flux analysis, the results presented 
in Table~\ref{table_f} assume that the total neutrino flux of neutrinos to be
detected at HELLAZ and Borexino is the one predicted by the SSM, {\it i.e.,}\/ there
is no ``room'' for other, yet unknown, low-energy solar neutrino sources. For this
reason, of course, the bounds obtained are (in some cases) much more stringent.  
\begin{table}
\caption{95\% CL 
Upper limits on the flux of solar muon-type neutrinos and antineutrinos
when the data after 5 years of HELLAZ/Borexino running is consistent
with SSM predictions, assuming that the total $pp$ ($^7$Be) neutrino flux is 
the one predicted by the SSM, with 2\% (20\%) (inflated) uncertainty.}
\begin{center}
\begin{tabular}{|c|c|c|c|}
\hline
Experiment & $\Phi_{\nu_{\mu}}/\Phi_{SSM}$ & $\Phi_{\bar{\nu}_{\mu}}/\Phi_{SSM}$ &
$\Phi_{\bar{\nu}_{e}}/\Phi_{SSM}$ \\ \hline
HELLAZ & 0.14 & 0.14 & 0.13 \\ \hline
Borexino & 0.66 & 0.68 & 0.058 \\ \hline
\end{tabular}
\end{center}
\label{table_f}
\end{table}    
    
Finally, as argued before, we emphasise that fixing the value of the solar neutrino 
flux to its SSM value is a reasonable thing to do, especially for $pp$-neutrinos. 
In these ``exclusion analyses'' such a procedure is even more natural, especially 
if one keeps in mind that a theoretical hypothesis, {\it i.e.}\/, 
$\nu_e\leftrightarrow\nu_{\mu}$ oscillations plus the SSM computed values for the 
solar neutrino flux, has been ``confirmed experimentally.''

\section{Conclusions\label{sec:conclusions}}

In order to unambiguously solve the solar neutrino puzzle, and to establish
the oscillations of solar neutrinos (if they occur), 
clear ``smoking gun'' signatures are required. Such 
signatures include a large day-night effect,
anomalous seasonal variations, or an obvious distortion of 
the neutrino energy spectrum. Another unambiguous signature is 
a discrepancy between the number of
charged current and neutral current events at SNO, which can be viewed
as an ``appearance'' experiment of $\nu_{\mu,\tau}$.  However, SNO can look 
for this ``appearance'' signature only for $^{8}$B neutrinos with
$E_{\nu} \gtrsim 6.5$~MeV and hence similar studies for lower energy 
neutrinos such as $^{7}$Be and $pp$ neutrinos, which are less sensitive to 
details of the solar model, are important.

We have argued in this paper that a careful analysis of the recoil kinetic
energy spectrum at Borexino and HELLAZ serves as another ``smoking gun'' 
signature, in the sense that one may be able to infer, independent of the SSM
prediction for the solar neutrino flux,
the existence of $\nu_{\mu,\tau}$ coming from the Sun. It is worthwhile
to emphasise that this is {\sl different}\/ from distortions in the incoming
neutrino energy spectrum. In our case we are describing an ``appearance''
experiment, while the analysis of the neutrino energy spectrum is a 
(energy dependent) ``disappearance'' experiment.

It is important to point out that, in our simple simulations, no background
events were included. While this is probably an oversimplification in the case
of Borexino, it may well be a good approximation for HELLAZ.  Moreover future 
upgrades of Borexino may reduce the background further according to 
recent encouraging progress \cite{1E-17}.  One should keep in
mind that, even if the background rates are significant, the procedure we 
described may still be useful if the background can be successfully dealt with
(one should not underestimate the ability and creativity of experimental 
physicists!). 

We have also included in the analysis the SSM prediction of the flux of solar
neutrinos.  While the results obtained
in this manner are model-dependent (they are not ``smoking gun'' signatures of
neutrino oscillations), we found them very useful.  This is a
reasonable thing to do especially for $pp$-neutrinos, whose flux is
constrained well by the solar luminosity.  This additional input makes
the measurement of the oscillation probability more precise.

Finally, we have argued that, {\sl if the data collected at Borexino and HELLAZ is
consistent with $\nu_{e}\leftrightarrow\nu_{\mu,\tau}$ oscillations},\/ one can 
try to exclude other neutrino oscillation modes ($\nu_{e}\leftrightarrow\nu_{s}$
and $\nu_{e}\leftrightarrow\bar\nu_{e,\mu,\tau}$) using the same procedure or, at
least, to set upper limits on the flux of solar antineutrinos. Again
we considered the possibility of constraining the solar neutrino flux to the 
SSM predicted value. The main result we obtained is that 
$\nu_{e}\leftrightarrow\bar\nu_{e}$ oscillations can, in general, be excluded,
while the $\nu_{e}\leftrightarrow\bar\nu_{\mu,\tau}$ case is much more elusive.
Nonetheless, Borexino should be able to exclude 
$\nu_{e}\leftrightarrow\bar\nu_{\mu,\tau}$ oscillations if the SMA MSW solution 
to the solar neutrino puzzle happens to be the correct one.

\section*{Acknowledgements} 
AdG thanks Hiroshi Nunokawa for questioning whether or not the issue of 
antineutrinos in the solar flux could be addressed by this procedure 
during a seminar at the Instituto de F\'{\i}sica Gleb Wataghin, 
UNICAMP, Brazil. The work of HM was supported in part by the Director, 
Office of Science, Office of High Energy and Nuclear Physics, Division of 
High Energy Physics of the U.S. Department of Energy under Contract  
DE-AC03-76SF00098 and in part by the National Science Foundation
under grant PHY-95-14797 and also by the Alfred P. Sloan Foundation.

\end{document}